\documentstyle[preprint,aps]{revtex}
\begin{document}
\tightenlines
\newcommand{\beq}{\begin{equation}}
\newcommand{\eeq}{\end{equation}}
\newcommand{\half}{{1\over 2}}
\newcommand{\fourth}{{1\over 4}}
\newcommand{\emath}{\mathrm{e}}
\newcommand{\contint}{{1\over{2\pi\imath}}\oint}
\newcommand{\iO}{{\it O}}
\newcommand{\lrarrow}[1]{\raisebox{1.4ex}{\makebox[0pt][l]
{$\leftrightarrow$}}#1}
\newcommand{\eqref}[1]{(\ref{#1})}
\newcommand{\V}[1]{V^{(#1)}}

\draft
\preprint{YITP-97-58}
\title{Commuting Charges of the Quantum Korteweg-deVries and Boussinesq 
Theories from the Reduction of $W_{\infty}$ and $W_{1+\infty}$ Algebras}
\author{Andrew J. Bordner\thanks{e-mail address: 
bordner@yukawa.kyoto-u.ac.jp} \\
Yukawa Institute for Theoretical Physics, Kyoto University, Kyoto 606, Japan}
\date{\today}
\maketitle
\begin{abstract}
Integrability of the quantum Boussinesq equation for $c=-2$ is
demonstrated by giving a recursive algorithm for generating explicit
expressions for the infinite number of commuting charges based on a
reduction of the $W_{\infty}$ algebra.  These
charges exist for all spins $s \geq 2$.  Likewise, reductions of the 
$W_{\infty/2}$ and $W_{(1+\infty)/2}$ algebras yield the commuting quantum
charges for the quantum KdV equation at $c=-2$ and $c=\half$,
respectively.  

\end{abstract}
\section{Introduction}
Both the classical KdV and Boussinesq equations are 
integrable Hamiltonian systems, {\it i.e.}, there are an
infinite number of integrals of local polynomials in the fields and their
derivatives, or charges, whose
Poisson brackets with both the Hamiltonian and each other 
vanish \cite{classical_Boussinesq}.  Of course, this implies an
infinite hierarchy of equations, however we refer only to the 
lowest-dimensional member.     
The algebra of fields induced by the second Poisson bracket for the KdV and
Boussinesq equations give, respectively, the Virasoro and classical
$W_{3}$ algebra.  We will study quantum versions of the
KdV and Boussinesq theories, which share the classical theories' integrability.

Sasaki and Yamanaka first considered the quantum KdV 
equation, in which the second Poisson brackets are replaced by the
commutator as $\imath\{,\}\rightarrow [,]$ and classical Hamiltonian
$\int u^{2}(x)dx$ is replaced by $\contint (TT)(z) dz$, where the
parentheses denote a normal-ordering of the quantum fields, to be
defined below \cite{Sasaki}.  They
found the existence of quantum charges $H_{s}$ of spin $s$, one for
each odd spin $1 \leq s \leq 9$, and postulated that charges 
exist for each odd spin. They also noticed that these conserved charges 
took a particularly simple form when the central charge, $c=-2$. 

The quantum Boussinesq theory has fields whose commutators are those
of the $W_{3}$ extended conformal algebra and likewise has a
integrable classical
limit as the central charge $c\rightarrow\infty$ and the commutators
are replaced with Poisson brackets. 
The equations of motion for the fields $T(z)$ and $W(z)$ in this
theory are
\begin{eqnarray}
\dot{T}&=&[T,P], \\
\dot{W}&=&[W,P], \\
P &=& \contint W dz .
\end{eqnarray}
The dot indicates the derivative with respect the ``time'' variable,
actually $\bar{z}$.  (Note that the field theory is formulated
using light-cone quantization and that the fields $T$ and $W$ depend 
on both $z$
and $\bar{z}$, however we drop the $\bar{z}$ dependence for notational 
simplicity.)
The fields $T$ and $W$ are the currents of the $W_3$ extended
conformal 
algebra with central charge $c$ and with operator product expansions
(OPEs) \cite{Zamolodchikov}
\begin{eqnarray}
\label{TT_OPE}
T(y)T(z) &=& {c/2\over(y-z)^4} + {2T(z)\over(y-z)^2} + {\partial T(z)
\over(y-z)} + \iO(1), \\
\label{TW_OPE}
T(y)W(z) &=& {3W(z)\over(y-z)^2}+{\partial W(z)\over(y-z)} + \iO(1),\\
\label{WW_OPE}
W(y)W(z) &=& {c/3\over(y-z)^6}+{2T(z)\over(y-z)^4}+{\partial T(z)
\over(y-z)^3}+{1\over(y-z)^2}\left[{3\over 10}\partial^2 T(z)  
+{32\over 22+5c}\Lambda(z)\right] \\ \nonumber
&+&{1\over(y-z)}\left[{1\over 15}\partial^3 T(z) 
+ {16\over 22+5c}\partial\Lambda(z)\right]+ \iO(1)
\end{eqnarray}
with 
\beq
\Lambda(z)=(TT)(z)-{3\over 10}\partial^{2}T(z).
\eeq
Parentheses around operators denote normal 
ordering defined by
\beq
(AB)(z) = \contint {dx\over x-z}A(x)B(z).
\eeq
Assuming spatial 
periodicity, Fourier modes for the fields may be introduced, which then 
gives the usual commutation relations for $W_{3}$ modes.

Kupershmidt and Mathieu considered the quantum Boussinesq theory for
generic central charge and found quantum charges for the first few lowest
spins $s$ for $s \neq 0 \bmod 3$ \cite{Kupershmidt_Mathieu}.  The
classical Boussinesq equation also lacks charges for these values of
the spin.

The quantum charges for the quantum KdV and Boussinesq theories may be 
related, through a quantum Miura transformation, to the charges of
$A_{1}$ and $A_{2}$ Toda field theory, respectively.
Quantum Toda field theories have been proven to be integrable, {\it
i.e.}, there exist an infinite number of mutually commuting charges which
also commute with the Hamiltonian \cite{Feigin}.  However, this proof
is not constructive in the sense that explicit forms for the conserved 
charges are not given.  

In this paper the commuting charges for the quantum Boussinesq theory
for $c=-2$ are constructed using the reduction of the 
$W_{\infty}$ algebra to
normal-ordered products of $W_{3}$ currents as described in
Ref. \cite{Lu}. 
We find similar
reductions 
of the infinite-dimensional 
algebras $W_{\infty/2}$ and $W_{(1+\infty)/2}$ which allow a 
construction of the commuting charges for the
quantum KdV theory at $c=-2$ and $c=1/2$.  Finally, the 
construction of the commuting conserved charges for
the associated $A_{1}$ and $A_{2}$ Toda field theories are 
briefly described. 

\section{Quantum conserved charges from $W_{\infty}$}
First the construction of $W_{N}$ from
$W_{\infty}$ at $c=-2$ given in Ref. \cite{Lu} is reviewed.   The
authors used a 
representation of the $W_{\infty}$ currents $\V{i}$, with spin $i+2$, 
for $c=-2$ in terms of fermionic ghost fields $b(z)$ and $c(z)$ as
\beq
\label{fermion_rep}
\V{i}(z) = \sum_{j=1}^{i+1} a_{j}(i)\partial^{j}c(z)
\partial^{i+1-j}b(z)
\eeq
with $a_{j}(i)$ the constants 
\beq
\label{a_def}
a_{j}(i)=\left(\begin{array}{c}i+1\\j\end{array}\right)
{(-1)^{i-j+1}(4q)^{i}(i+3-j)_{j}(j)_{i+1-j}
\over (i+2)_{i+1}}
\eeq
and $(a)_{n}\equiv \Gamma(a+n)/\Gamma(a)$ \cite{fermion_rep}. 
The only non-zero OPE of the fermion fields is $b(z)c(w)\sim
1/(z-w)$.  $q$ is an
arbitrary parameter which fixes the normalization of the 
currents; $q^{2}=1/16$ in
Ref. \cite{Lu} however throughout this paper $q^{2}=1/24$ is chosen to 
agree with the
normalization of Eqs. \ref{TT_OPE}-\ref{WW_OPE} with the 
definitions $T(z)\equiv
\V{0}(z)$ and $W(z)\equiv \V{1}(z)$.  By requiring the 
vanishing of quartic fermion
terms resulting from the normal-ordered product of two $W_{\infty}$
currents in the fermion representation Eq. \ref{fermion_rep}, it was shown 
that the $W_{\infty}$ current $\V{i}$ for $i \geq
2$ may be expressed in terms of only currents of lower spin as 
\beq
\label{lower_spin_soln}
\V{i}= \beta_{i}\sum_{p=0}^{i-2}b_{p}^{(i)}(\V{p}\V{i-2-p}) 
+ \sum_{p=1}^{[{i\over 2}]}\gamma_{2p}^{(i)}\partial^{2p}\V{i-2p},
\eeq
where $[x]$ denotes the integer part of $x$.
The coefficients $\beta_{i}$, $b_{p}^{(i)}$, and 
$\gamma_{p}^{(i)}$ are
\begin{eqnarray}
\beta_{i}&=&{c_{i}\over S_{i}(i)}, \\
b_{p}^{(i)} &=& {a_{i-1-p}(i-2)\over 2a_{1}(p)a_{i-1-p}(i-2-p)}, \\
\gamma^{(i)}_{p} &=&
-{\beta_{i}S_{i-p}(i)(2i+3-2p)!\over(2i+3-p)!c_{i-p}}.
\end{eqnarray} 
$c_{i}$ is defined by 
\beq
\label{c_i}
c_{i}={(2q)^{2i}i!(i+2)!(2i+2)!!\over8(2i+1)!!}c
\eeq
with the central charge $c=-2$ and 
\begin{eqnarray}
S_{j}(i)
&=&\sum_{p=0}^{i-2}\sum_{k=1}^{p+1}\sum_{l=1}^{i-1-p}\sum_{m=1}^{j+1}b_{p}^{(i)}(-1)^{k+j}a_{k}(p)a_{l}(i-2-p)a_{m}(j) 
\\ \nonumber
&\times&\left({(i-l+1+m)!(j+1+l-m)!\over 
(i-p+k-l)}\right. \\ \nonumber
&+&\left.(-1)^{p}{(i-p-1-l+m)!(p+l+3+j-m)!\over(p-k+l+2)}\right).
\end{eqnarray}

For example, the first few lowest spin currents may be written as 
\begin{eqnarray}
\label{lowest_spin}
\V{2} &=& {2\over 3}\left[(\V{0}\V{0})-{3\over
10}\partial^{2}\V{0}\right], \\ \nonumber
\V{3} &=& {2\over 5}\left[(\V{0}\V{1})+(\V{1}\V{0})-{13\over
14}\partial^{2}\V{1}\right], \\ \nonumber
\V{4} &=& {4\over
93}\left[5(\V{0}\V{2})+12(\V{1}\V{1})+5(\V{2}\V{0})-{40\over 
3}\partial^{2}\V{2}-{10\over 21}\partial^{4}\V{0}\right], \\
\nonumber
\V{5} &=& {5\over 168}\biggl[{7\over
2}(\V{0}\V{3})+15(\V{1}\V{2})+(\V{2}\V{1})+{7\over
2}(\V{3}\V{0})-{1203\over 44}\partial^{2}\V{3} \\
\nonumber
&-&{265\over
126}\partial^{4}\V{1}\biggr], \\ \nonumber
\V{6} &=& {2\over
907}\biggl[21(\V{0}\V{4})+140(\V{1}\V{3})+250(\V{2}\V{2})+140(\V{3}\V{1}) \\ \nonumber
&+&21(\V{4}\V{0})-{6454\over 
13}\partial^{2}\V{4}-{6005\over 99}\partial^{4}\V{2}-{103\over 81}
\partial^{6}\V{0}\biggr].
\end{eqnarray}

If one defines the mode expansion for $W_{\infty}$ currents as 
\beq
\V{i}(z) \equiv \sum_{m}\V{i}_{m}z^{-m-i-2}
\eeq
the commutation relations for the modes then have the standard form for Virasoro quasi-primary fields \cite{quasi-primary}
\beq
\label{mode_comm}
[\V{i}_{m},\V{j}_{n}]=
\sum_{k =
0}^{i+j} C^{k}_{ij}p_{ijk}(m,n)V^{k}_{m+n}+\delta_{n,-m}D_{ij}\left(\begin{array}{c}m+i+1\\2i+3 \end{array}\right)
\eeq
where $C_{ij}^{k}=0$ for $i+j-k$ an odd number.
$p_{ijk}(m,n)$ are the universal polynomials
\beq
p_{ijk}(m,n) = \sum_{r,s = 0}^{\Delta_{i}+\Delta_{j}-\Delta_{k}-1}\delta_{r+s,\Delta_{i}+\Delta_{j}-\Delta_{k}-1}c^{ijk}_{r,s}\left(
\begin{array}{c}m+\Delta_{i}-1\\ r\end{array}\right)\left(\begin{array}{c}
n+\Delta_{j}-1\\s\end{array}\right)
\eeq
with
\beq
c^{ijk}_{r,s}=(-1)^{s}{(2\Delta_{k}-1)!\over (\Delta_{i}+\Delta_{j}+\Delta_{k}-2)!}(-\Delta_{i}+\Delta_{j}+\Delta_{k})_{r}(\Delta_{i}-\Delta_{j}+\Delta_{k})_{s}.
\eeq
$\Delta_{i} = i+2$ is the conformal dimension of the field $\V{i}$.
The structure constants are \cite{W_infty_comm} 
\beq
C_{ij}^{k}=q^{i+j-k}{(i+j-k+4)!\over 2(2k+3)!}\sum_{r=0}{{(-\half)_{r}({3\over
2})_{r}
(-\half(i+j-k+1))_{r}(-\half(i+j-k))_{r}}\over{r!(-i-\half)_{r}(-j-\half)_{r}(k+{5\over
2})_{r}}}
\eeq
for $i+j-k$ even and $0$ otherwise.
The central term is $D_{ij}= \delta_{ij}c_{i}$ with 
$c_{i}$ defined in Eq. \ref{c_i}.  

The $W_{\infty}$ algebra has an infinite-dimensional Abelian
subalgebra consisting of the modes $\V{i}_{-i-1}$.  These are the 
commuting charges of the quantum
Boussinesq equation.  The charges $\V{i}_{-i-1}$ in the $c=-2$
fermion representation, Eq. \ref{fermion_rep}, are those used in
Ref. \cite{DiFrancesco} to construct charges of the quantum KdV
equation.  They are 
\begin{eqnarray}
\label{Abelian_subalgebra}
H_{j+1} &=& \contint \V{j} dz = \V{j}_{-j-1} \\ \nonumber
&\sim& \contint (\partial^{j+1}c(z)b(z)) dz
\end{eqnarray}
with the tilde denoting proportionality up to an (inessential)
constant.
In fact the modes $\V{i}_{-i-1}$ of the dimension $i+2$ currents of any Lie
algebra of Virasoro quasi-primary fields form an Abelian subalgebra if 
the structure constants $C_{ij}^{i+j+1}$ vanish.

The procedure to generate the quantum conserved charges for the Boussinesq 
theory is similar to the construction of the $c=-2$ $W_{3}$ currents 
in terms of $W_{\infty}$ currents described in Ref. \cite{Lu}.  First 
the $W_{\infty}$ 
currents $T \equiv \V{0}$ and $W \equiv \V{1}$ are regarded as
fundamental 
and the 
currents with higher spin are replaced by their expressions quadratic in 
lower spin currents using Eq. \ref{lower_spin_soln}.  This is a recursive 
procedure which results in a representation of the $W_{\infty}$ algebra 
by normal-ordered products of $W_{3}$ currents.  The commuting charges 
are then the contour integral of 
these reduced $W_{\infty}$ currents as given in Eq. \ref{Abelian_subalgebra}.
This procedure readily yields the 
charges for quantum Boussinesq theory as follows    
\begin{eqnarray}
\label{Boussinesq_charges}
H_{1} &=& \contint T(z) dz, \\ \nonumber
H_{2} &=& \contint W(z) dz, \\ \nonumber
H_{3} &=& \contint (TT)(z) dz, \\ \nonumber
H_{4} &=& \contint (TW)(z) dz, \\ \nonumber
H_{5} &=& \contint \left[(WW)(z)+{5\over 9}(T(TT))(z)
+{1\over 6}(\partial T\partial T)(z)\right] dz, \\ \nonumber
H_{6} &=& \contint \left[(\partial T\partial
W)(z)+2(W(TT))(z) \right] dz, \\ \nonumber
H_{7} &=& \contint \biggl[(\partial W\partial W)(z)-{119\over 31}(T(WW))(z)
-{4295\over 2232}(T(T(TT)))(z)+{6139\over
8928}(\partial^{2}T\partial^{2}T)(z)\\ \nonumber
&-&{4949\over
2232}(\partial^{2}T(TT))(z)\biggr] dz.
\end{eqnarray} 
These expressions have been multiplied by constants for convenience since 
this does not affect their commutativity.  
The following equation for the normal-ordered commutator
\beq 
([A,B])(z)  = \sum_{n>0}{(-1)^{n+1}\over n!}\partial^{n}\{AB\}_{n}(z)
\eeq 
in terms of the operators $\{AB\}_{n}(z)$ appearing in the OPE of $A(y)$
and $B(z)$ 
\beq
A(y)B(z) = \sum_{n=-\infty}^{M}{\{AB\}_{n}(z)\over (y-z)^{n}},
\eeq
with $M$ is a positive integer, as well as the rearrangement relation
\beq
(A(BC))(z)-(B(AC))(z)=(([A,B])C)(z)
\eeq 
have been used to simplify the resulting expressions.
Unlike for generic values of the central charge, one finds that
charges exist for all
integer spins $s\geq 1$ for $c=-2$.

$W_{N}$ minimal models, whose Verma modules contain an infinite number 
of singular vectors, exist for certain values of the central 
charge $c$ specified by two relatively prime integers $p$ and $p'$
according to
\beq
c = N-1-{(p-p')^{2}\over pp'}N(N^2-1).
\eeq
In the case of $c=-2$ $W_{N}$ minimal models exist for all $N$ ($p'=N-1$,
$p=N$).  Consider the conservation of the charge $H_{3}$.  The
commutator depends only on the simple pole term and gives
\begin{eqnarray}
\dot{H}_{3} &=& \left[\contint dz (TT)(z),\contint dy W(y)\right] \\
&=& \contint dz {1\over
5}\left(-12\partial(TW)(z) - \partial^{3}W(z)+\Xi(z)\right) 
\end{eqnarray}
with $\Xi(z) \equiv -8(T\partial W)(z)+12(\partial
TW)(z)+\partial^{3}W(z)$.  The total derivative terms as well as
$\Xi(z)$ vanish resulting in a conserved $H_{3}$.
$\Xi(z)$ vanishes since it is a null descendent of the Virasoro
primary field $W(z)$ at level $3$.  In other words, defining
\beq
\hat{L}_{-n}\phi(w) \equiv \contint {dz\over (z-w)^{n-1}} T(z)\phi(w)
\eeq
then $\Xi(z) =
(-8\hat{L}_{-2}\hat{L}_{-1}+12\hat{L}_{-3}+\hat{L}_{-1}^{3})W(z) =
0$.  One may show that $\Xi(z)$, and all other singular vectors, vanish 
identically in the fermion representation since all products of modes of 
fermion fields with equal number of modes of $b(z)$ and $c(z)$ may be 
expressed in terms of modes of $T(z)$ and $W(z)$.  The singular vectors are 
then states in the nonsingular Fock space with zero inner product with all 
other states and thus are the unique state with zero norm in the 
fermion Fock space.

\section{Reduction of $W_{\infty/2}$}
The $W_{\infty}$ algebra contains an infinite-dimensional subalgebra
$W_{\infty/2}$ consisting of only the even spin currents of
$W_{\infty}$.  One may see from Eq. \ref{lowest_spin} that a different set of
normal-ordered products of lower-dimensional currents is necessary for 
the reduction of $W_{\infty}/2$, however the central charge is the same 
as in the $W_{\infty}$ reduction, namely $c=-2$.  The general form of 
the reduction is 
\beq
\V{2i} =
\beta_{i}\sum_{p=0}^{i-1}b_{p}^{(2i)}(T\partial^{2p}\V{2i-2p-2})
+\sum_{p=1}^{i}\gamma_{2p}^{(2i)}\partial^{2p}\V{2i-2p}.
\eeq
Although there is no simple expression for the coefficients they may
be found in the same manner as for the reduction of $W_{\infty}$.
The coefficients $b_{p}^{(2i)}$ are determined by the vanishing of
quadratic fermion terms in the sum of normal-ordered products of
currents.  This leads to a linear system of equations
\beq
\sum_{p=0}^{i-1}b_{p}^{(2i)}\sum_{k=0}^{\min(2m,l-1)}\left(\begin{array}{c}2m 
\\ k \end{array}\right)a_{l-k}(2i-2p-2)=0,\qquad l=2,\ldots,i
\eeq
with $a_{j}(i)$ given by Eq. \ref{a_def}.  This has a unique solution 
up to an overall multiplicative factor which may be absorbed in the
coefficient $\beta_{i}$.  The remaining coefficients may then be found 
by considering the central term in the operator product expansion of
two currents as explained in Ref. \cite{Lu}.  The reduction of
$W_{\infty/2}$ for the lowest spin currents is then  
\begin{eqnarray}
\V{2} &=& {2\over 3}\left[(TT)-{3\over
10}\partial^{2}T\right], \\ \nonumber
\V{4} &=& {8\over 9}\left[(T \V{2})+{1\over
5}(T\partial^{2}T)\right]-{2\over 9}\partial^{2}\V{2}
-{32\over 945}\partial^{4}T, \\ \nonumber
\V{6} &=& (T\V{4})+{10\over
9}(T\partial^{2}\V{2})
+{4\over 63}(T\partial^{4}T)-{9\over 13}\partial^{2}\V{4}-{5\over 33}\partial^{4}\V{2}\\ \nonumber 
&-&{5\over
567}\partial^{6}T, \\ \nonumber
\V{8} &=& {16\over 15}\left[(T\V{6})+{35\over 13}(T\partial^{2}\V{4})
+{70\over 99}(T\partial^{4}\V{2})+{2\over 81}(T\partial^{6}T)\right] \\ \nonumber
&-&{148\over 85}\partial^{2}\V{6}-{56\over
117}\partial^{4}\V{4}-{4592\over 57915}\partial^{6}\V{2}-{64\over 22275}\partial^{8}T, \\ \nonumber
\V{10} &=& {10\over 9}\biggl[(T\V{8})+{84\over
17}(T\partial^{2}\V{6})
+{112\over 39}(T\partial^{4}\V{4})
+{1568\over 3861}(T\partial^{6}\V{2}) \\ \nonumber
&+&{16\over 1485}(T\partial^{8}T)\biggr]
-{215\over 63}\partial^{2}\V{8}
-{4480\over 2907}\partial^{4}\V{6}-{1960\over 5967}\partial^{6}\V{4}\\ \nonumber
&-&{1360\over 34749}\partial^{8}\V{2}
-{112\over
104247}\partial^{10}T.
\end{eqnarray}
Following the same procedure as in the preceding section, one then 
obtains the quantum commuting charges for the quantum KdV at $c=-2$.
These are the charges found previously in Ref. \cite{DiFrancesco}, namely 
$H_{2n-1} = \contint(\ldots(((TT)T)T)\ldots T)(z)dz$ with $n$ factors of T.

\section{Reduction of $W_{(1+\infty)/2}$}
$W_{1+\infty}$ is another infinite-dimensional Lie algebra generated
by fields quasi-primary with respect to $T \equiv \V{0}$ of spins $i
\geq 1$ \cite{one_plus_infty}.  One finds a reduction of this algebra 
for $c=\half$.  
The commutation relations for the modes 
are given by Eq. \ref{mode_comm} with structure constants
\beq
C_{ij}^{k}=q^{i+j-k}{(i+j-k+4)!\over 2(2k+3)!}\sum_{r=0}{{(\half)_{r}({1\over
2})_{r}
(-\half(i+j-k+1))_{r}(-\half(i+j-k))_{r}}\over{r!(-i-\half)_{r}(-j-\half)_{r}(k+{5\over
2})_{r}}}
\eeq
and central terms
\beq
D_{ij} = \delta_{ij}{{(2q)^{2i}((i+1)!)^{2}(2i+2)!!}\over {4(2i+1)!!}}c.
\eeq
Since we would like to consider only currents of conformal dimension
$\Delta_{i}\geq2$ we restrict to the subalgebra $W_{(1+\infty)/2}$ of
even-dimensional currents.  We find a representation of this
subalgebra at central charge $c=\half$ by a real fermion $\psi$ as
\beq
\V{2i}(z) = {2^{2i-1}(2i+1)!\over
(4i+1)!!}q^{2i}\sum_{k=0}^{i}(-1)^{k}\left(\begin{array}{c}2i+1 \\ k
\end{array}\right):\partial^{2i+1-k}\psi(z)\partial^{k}\psi(z): 
\eeq
with $\psi(z)\psi(w)\sim 1/(z-w)$.  Note that this is not the 
usual representation of $W_{1+\infty}$ by complex
fermions, which has central charge $c=1$ \cite{one_plus_infty_rep}.  
Again, using the methods of Ref. \cite{Lu} and a reduction of the form 
in Eq. \ref{lower_spin_soln} with a summation only over even-spin
currents, we obtain the following reduction for the lowest spin currents:
\begin{eqnarray}
\V{2} &=& {4\over 7}\left[(TT)-{3\over 10}\partial^{2}T\right], \\ \nonumber
\V{4} &=& {400\over 441}\left[(T\V{2})-{1\over
6}\partial^{2}\V{2}\right], \\ \nonumber
\V{6} &=& {1323\over
5743}\biggl[(T\V{4})+{2500\over 567}(\V{2}\V{2})-{2581\over
702}\partial^{2}\V{4}-{2375\over
8019}\partial^{4}\V{2} \\ \nonumber
&-&{50\over 15309}\partial^{6}T\biggr], \\ \nonumber
\V{8} &=& {3744\over 87545}\biggl[(T\V{6})+{5145\over
143}(\V{2}\V{4})-{54237\over
2431}\partial^{2}\V{6}-{117649\over
66924}\partial^{4}\V{4} \\ \nonumber
&-&{85750\over
4969107}\partial^{6}\V{2}\biggr].
\end{eqnarray}
The quantum commuting charges for $c=\half$ quantum KdV may be found from this
reduction using the same method as for the reduction of the
$W_{\infty}$ algebra.  The corresponding charges are: 
\begin{eqnarray}
H_{3} &=& \contint (TT)(z)dz \\ \nonumber
H_{5} &=& \contint \left[(T(TT))(z)+{3\over 10}(\partial T\partial
T)(z)\right]dz \\ \nonumber
H_{7} &=& \contint \left[(T(T(TT)))(z)+{77\over
85}(\partial^{2}T(TT))(z)+{809\over 4420}(\partial^{2}
T\partial^{2}T)(z)\right] dz \\ \nonumber
H_{9} &=& \contint \biggl[(T(T(T(TT))))(z)+{1089043\over
897915}(\partial^{2}T(T(TT)))(z)-{998002\over 179583}(\partial
T(\partial T(TT)))(z) \\ \nonumber
&+&{22912073\over
86199840}(\partial^{4}T(TT))(z)+{1414087427\over 244232880}(\partial^{2}T(\partial
T\partial T))(z)+{57504805453\over
571504939200}(\partial^{3}T\partial^{3}T)(z)\biggr] dz.
\end{eqnarray}
\section{Commuting Charges for $A_{1}$ and $A_{2}$ Toda Field
Theories}
The quantum commuting charges found above may be used to construct
charges for $A_{1}$ and $A_{2}$ Toda field theories via
the quantum Miura transformation.  However these charges only commute
modulo singular vectors of the corresponding Virasoro or $W_{3}$ algebras,
respectively.  The holomorphic part of the
imaginary coupling $A_{N}$ 
Toda
field theory interaction Hamiltonian, in light-cone coordinates, is
\beq
\label{Toda_H}
V_{0} \equiv \sum_{i=1}^{N}\contint dz :\exp\left(-\imath\beta\alpha_{i}
\cdot\phi(z)\right):.
\eeq
$(\alpha_{i}$, $i=1,\ldots,N)$ is a basis of simple roots for $A_{N}$, 
which we may normalize to $|\alpha_{i}|^{2}=2$, $\beta$ is a real
coupling constant, and $\phi$ is a set of $N$ hermitian scalar fields.
This theory is a conformal field theory with central charge
$c=N\left[1-(N+1)(N+2)(2\beta-1/\beta)^{2}\right]$ \cite{Toda_conformal}. 

The algebra of the
quantum charges for $A_{N}$ Toda field theory is the $W_{N+1}$
algebra whose currents are found from the quantum Miura
transformation.  This has the following generating expression:
\beq
-\sum_{k=0}^{N+1}U_{k}(z)(\gamma_{0}\partial)^{N+1-k}=((\gamma_{0}\partial-\epsilon_{1}\cdot\imath\partial\phi(z))\cdots(\gamma_{0}\partial-\epsilon_{N+1}\cdot\imath\partial\phi(z)))
\eeq
where $(\epsilon_{i}, i = 1,\ldots,N+1)$ are the weights of the vector 
representation of $A_{N}$ with
$\alpha_{i}=\epsilon_{i}-\epsilon_{i+1}, i=1,\ldots,N$,
$\epsilon_{i}\cdot\epsilon_{j}=\delta_{ij}-1/(N+1)$, and
$\sum\epsilon=0$ \cite{quantum_Miura}.  $\gamma_{0}$ is a constant 
related to the central charge by $c=N(1-(N+1)(N+2)\gamma_{0}^{2})$.  Then $T=U_{2}$, $W=U_{3}-\half(N-1)\gamma_{0}\partial U_{2}$, and the higher 
spin currents
of the $W_{N}$ algebra are found by projecting to Virasoro
quasi-primary combinations of the $U_{i}$ \cite{primary_proj}.  
The $U_{i}(z)$, and hence any products of modes of them, commute with
each term in the Toda Hamiltonian Eq. \ref{Toda_H}.  These terms in
the potential are
generalization of the screening currents of the Feigin-Fuchs
formulation of minimal conformal models, {\it i.e.}, they have 
conformal dimension one.  

Consider, more
specifically, $A_{2}$ Toda field theory.  Any sum of normal-ordered
products of $T$ and $W$ and their derivatives commute with the
Hamiltonian $V_{0}$.  The difficulty is to find a commuting set.  
Charges which commute, modulo singular vectors of the $W_{3}$
representation, are provided by
Eq. \ref{Boussinesq_charges} for $c=-2$.  Unlike in
the fermion representation of Eq. \ref{fermion_rep}, the singular
vectors do not vanish identically in the boson representation.  The
boson operators corresponding to these singular vectors should be set
equal to zero, which results in an irreducible representation of the
$W_{3}$ algebra.  Thus, one obtains an infinite set of charges which
commute modulo singular vectors for $A_{1}$ Toda field theory with 
couplings $\beta_{+}=(\sqrt{2}/8)(1+\sqrt{17})$ or 
$\beta_{+}=(\sqrt{3}/24)(1+\sqrt{97})$ and for the $A_{2}$ theory with 
coupling $\beta_{+}=\sqrt{2/3}$ using the preceding reductions.  Because 
of a coupling duality, these are 
also charges for the corresponding Toda field theories with coupling 
$\beta_{-}=-1/(2\beta_{+})$.

\section{Conclusion}

We have presented a single reduction of each of $W_{\infty}$,
$W_{\infty/2}$, and $W_{(1+\infty)/2}$ for given values of the central
charge.  One may show that these are the only reductions of
these algebras.  This is demonstrated by considering the general form of
the reduction of the spin $4$ current as $\V{2} =
b_{1}(TT)+\gamma_{2}\partial^{2}T$.  Then by requiring that
$\hat{L}_{1}\V{2}=0$ and that $D_{02}$, $D_{22}$, $C_{02}^{0}$, and
$C_{22}^{2}$ match those for the appropriate algebra, one finds only
the single solution, stated in the previous sections, for each algebra. 

Furthermore, one may also try to find other infinite-dimensional
Lie algebras of quasi-primary fields formed from modes of $W_{3}$ fields
$T$ and $W$.  However, by examining the commutators for the unique
spin $4$ field, one finds that there is only one algebra, the
$W_{\infty}$ algebra with $c=-2$, with exactly one quasi-primary field for each
conformal dimension $\Delta_{i}\geq 2$ generated by modes of $T$ and $W$.

There are several interesting questions to investigate further.   Since we 
have explicit expressions for commuting quantum
charges of the quantum Boussinesq and quantum KdV theories, one may
be able to find the common eigenvalues and eigenstates of these charges.  
In addition, it is not known whether there are similar infinite-dimensional 
Lie algebras of Virasoro quasi-primary fields, which likewise give the 
commuting quantum charges, for theories whose fields are currents of 
other finite-dimensional $W$ algebras.
Finally, since $W$ algebras for 
certain values of the 
central charge may be represented via the Kac-Moody algebra coset 
construction, one may find that the Abelian subalgebra of the quantum charges 
has a simpler form in that case \cite{coset}.

\section*{Acknowlegements}
We thank Ryu Sasaki for many helpful discussions.  The author was
supported by the National Science Foundation under grant no. 9703595
and the Japan Society for the Promotion of Science.

\end{document}